# MICROWAVE-ASSISTED SYNTHESIS AND CHARACTERIZATION OF PEROVSKITE OXIDES

*Jesús Prado-Gonjal[1], Rainer Schmidt[2] and Emilio Morán[1]*

[1]Dpto. Química Inorgánica I, Facultad CC. Químicas, Universidad Complutense, Madrid, Spain
[2]Dpto. Física Aplicada III, GFMC, Facultad CC. Físicas,
Universidad Complutense, Madrid, Spain

## ABSTRACT

The use of microwave irradiation is a promising alternative heat source for the synthesis of inorganic materials such as perovskite oxides. The method offers massive energy and time savings as compared to the traditional ceramic method. In this work we review the basic principles of the microwave heating mechanism based on interactions between dipoles in the material and the electromagnetic microwave.

Furthermore, we comment on and classify all different sub-types of microwave synthesis such as solid-state microwave and microwave assisted hydrothermal synthesis. In the experimental part of this work we present a large range of materials that were synthesized in our laboratories by one of the microwave techniques, where such materials include superconducting, ferromagnetic, ferroelectric, dielectric and multiferroic perovskite systems.

## 1. INTRODUCTION

Perovskite oxides are a fascinating class of inorganic materials [1], which are usually prepared by means of the conventional ceramic method. Such method is especially well suited for oxides since reactions are carried out in air at ambient pressure. It involves the homogeneous grinding and mixing of stoichiometric amounts of solid precursor oxide reactants followed by heating at elevated temperatures for long periods of time (typically 1000 ºC and several hours or days). The grinding and heating cycle is repeated until the desired pure phase is obtained. Although there can be no doubt about the usefulness and easiness of this universal method of synthesis which works in thermodynamic equilibrium conditions, there are some intrinsic drawbacks such as the need for high temperatures to accelerate diffusion -which prevents metastable phases to appear-, the volatility of some reactants, the need of regrinding several times or the high energy and time consumptions. On these grounds, there is a permanent stimulus for searching alternative routes such as decomposition of alternative precursors, sol-gel methods and many more, where the reaction times can often be notably reduced. Synthesis is achieved in out-of-equilibrium conditions and such methods may be called "Fast Chemistry"; among them combustion, sonochemistry, "spark plasma" and microwave-assisted methods of synthesis, where the latter is the focus of this work. Many perovskite oxides can be prepared by these means as will be demonstrated in the following.

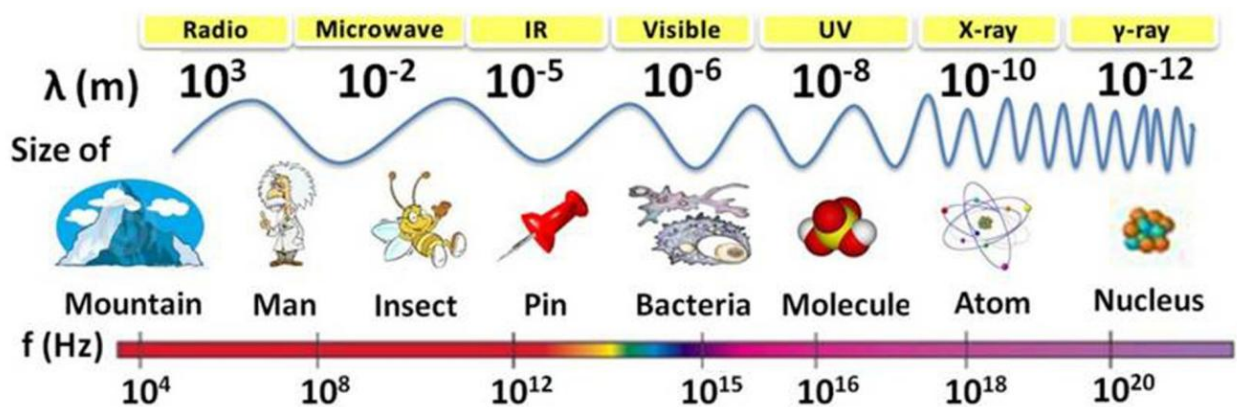

**Figure 1.** Electromagnetic spectra. Microwave frequencies range between 0.3 and 300 GHz.

Microwave research is indebted to British scientists J. Randall and H. A. Boot, who designed a magnetron valve to generate microwaves for radar systems back in 1940. Six years later, Percy Spencer, an engineer working on magnetrons for the Raytheon Company, accidentally discovered the rapid effect of microwaves on heating a chocolate bar and after some proofs being made with other kinds of food it was realized that microwave cooking was feasible and the path for a new kind of domestic furnaces, nowadays of general use, was open [2]. Nevertheless, it was not until 1975 that the heating effect of microwaves on ceramic materials was observed, described and studied by Sutton [3]. At a first glance, these heating effects are quite surprising since microwave radiation ranges in the low-energy zone of the electromagnetic spectrum (microwave frequencies range between 0.3 and 300 GHz), (Figure 1) [4]. Most microwave heating effects can be understood in terms of electric dipoles in a material which follow the alternating electric component of the electromagnetic microwave, i. e. dipolar molecules such as water do rotate, and the resistance to that movement generates considerable amount of heat [5].

For domestic purposes the conventionally used frequency is 2.45 GHz equivalent to 0.0016 eV, which is far below the energy of even a weak bonding energy such as the hydrogen bond (0.21 eV) [4].

Microwave assisted synthesis flourished firstly in the pharmaceutics and organic industry but in solid state chemistry and materials science it is still in its infancy; there are still many questions to be answered such as fundamental details on the reaction mechanisms and kinetic aspects of the reactions. The lack of such knowledge turns the method into a trial and error one at the current stage [6]. Therefore, the aims of this review article are:

i) To comment on fundamental aspects of the interactions between microwaves and solid matter.
ii) To compare conventional heating versus microwave heating.
iii) To demonstrate that microwave assisted synthesis is a fast and efficient method to produce perovskite oxide materials.
iv) To compare several ways to perform microwave assisted synthesis: solid state, single-mode and solvothermal routes.
v) To give relevant examples from our own work or taken from the literature of important perovskite materials in different areas such as energy materials (components for solid oxide fuel cells SOFC's), magnetic, ferroelectric, multiferroic materials and superconductors.

## 2. ON THE INTERACTION BETWEEN MICROWAVES AND MATTER

Conventional electromagnetic waves consist in 2 components of alternating magnetic and electric fields with perpendicular orientations. Both type of field osicalte in-phase with the same frequency and propagation in vacuum or air usually occurs with a constant amplitude. An electromagnetic microwave may be reflected, transmitted or absorbed by condensed matter depending on the nature of such matter. Therefore, in this context of condensed matter interacting with microwaves three classes of materials can be defined [7]:

i) *Reflecting materials,* which do not allow penetration of the electromagnetic wave; materials with free electrons such as certain metals behave in this way.

ii) *Transparent materials,* which allow the wave to propagate all the way through with low attenuation, such as Teflon and silica glass. Generally speaking, insulators or materials with low dielectric permittivity and low losses are concerned here.

iii) *Absorbing materials,* which transform the electromagnetic energy into heat; this is the case for dipolar liquids such as water and dielectric or polar materials such as ferroelectrics with high dielectric permittivity. The absorption of the microwave energy by the material is determined by the dielectric permittivity and the dielectric losses. An ideal microwave absorber has high dielectric permittivity and an intermediate level of losses.

The dielectric permittivity describes the "polarizability" of the material where the overall polarization is termed ($\alpha_1$). This term includes several accumulative components: electronic polarization ($\alpha_e$), atomic polarization ($\alpha_a$), dipolar polarization ($\alpha_p$), ionic polarization ($\alpha_{ion}$) and interfacial polarization ($\alpha_{if}$):

$(\alpha_1) = (\alpha_e) + (\alpha_a) + (\alpha_p) + (\alpha_{ion}) + (\alpha_{if})$

- Electronic polarization $\alpha_e$: when the atom is subjected to an external electric field, redistribution of the charges occurs and the atomic electron cloud shifts away from the equilibrium position with respect to the positive nuclei resulting in an induced dipole moment. This effect occurs in all solid matter containing localized electrons.
- Atomic polarization $\alpha_a$: This is equivalent to the electronic polarization, except that it accounts for the shift of the positively charged atomic cores in the opposite direction. It is usually small as compared to $\alpha_e$.
- Dipolar polarization $\alpha_p$: In thermal equilibrium, electric dipoles in the material may be randomly oriented and thus carry no net polarization. An external field can align these dipoles to some extent and thus induces a net polarization to the material. This contribution only exists in polar materials such as dielectrics with a strong ionic character of the chemical bonds.
- Ionic polarization (or ionic displacement) $\alpha_a$: This occurs due to relative displacements of positive and negative ions from their equilibrium lattice sites. The positively and negatively charged ions in the crystals are displaced spontaneously from their equilibrium lattice positions and the centers of negative and positive charge within one unit cell do

  not coincide. This constitutes a net dipole moment, which can be switched by high applied electric fields. This effect only occurs in piezo- and ferroelectrics.
- Interfacial polarization $\alpha_i$: Surfaces, grain boundaries, interphase boundaries (including the surface of precipitates) may be charged, i.e. they contain dipoles which may become oriented to some degree in an external field and thus contribute to the total polarization of the material. Such effects can occur in all kind of condensed matter, but the quantitative contribution to the total polarization is usually small due to the generally small volume fraction of interfaces in bulk material.

Polarization highly depends on the frequency of the applied electric field as illustrated in Figure 2. At low frequencies all dipoles can follow the applied field, but with increasing frequency the different contributions to the overall polarization consecutively "relax out", which involves that the respective dipoles cannot follow the alternating applied electric field anymore and do not polarize. This is reflected in a consecutive reduction of the dielectric permittivity $\varepsilon'$. Atomic and electronic contributions show a double-peak structure in the dielectric permittivity, which occurs when the frequency of the applied field matches the resonance frequency of the switching dipoles.

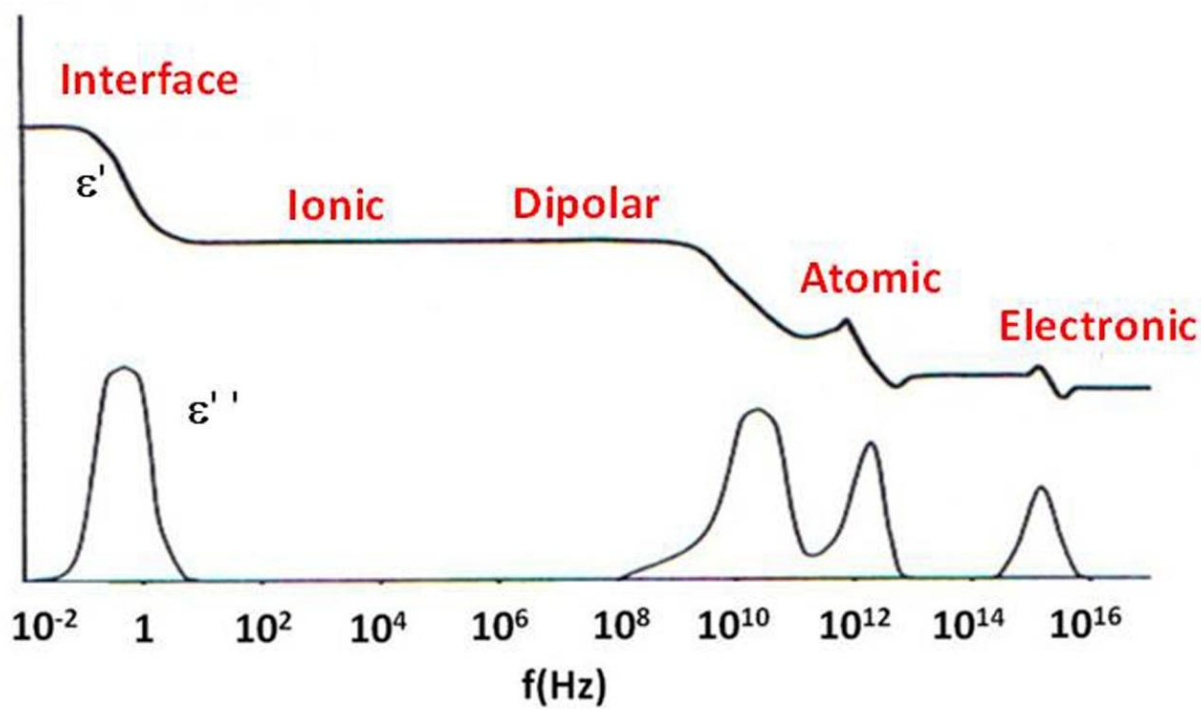

**Figure 2.** Frequency dependence of the polarization mechanisms (image reproduced from the web: http://en.wikipedia.org/wiki/Dielectric_spectroscopy).

In materials with high dielectric losses the microwave gets attenuated. Such attenuation can be improved by the addition of small amounts of metallic powder, which reflects the microwave and hinders loss-free propagation of the wave. The power density of absorbed microwaves may be expressed as:

$$P (W/m^3) = \text{electrical losses} + \text{magnetic losses} = \omega \, \varepsilon_o \, \varepsilon'' \, E^2_{rms} + \omega \, \mu_o \, \mu'' \, H^2_{rms}$$

where $\omega = 2\pi f$ = angular frequency, $\varepsilon_o$ = dielectric permittivity of vacuum, $\varepsilon''$ = relative dielectric losses (polarization + conduction), $E_{rms}$ = internal electrical field of the microwave, $\mu_o$ = magnetic permittivity in vacuum, $\mu''$ = relative magnetic losses and $H_{rms}$ = internal magnetic field of the microwave. The magnetic losses μ" are usually less relevant and the susceptibility of a polar material to microwave heating is mainly determined by the dielectric losses.

In polar materials with losses $\varepsilon" < 10^{-2}$ are quite difficult to be heated up, while for those with $\varepsilon"_{eff} > 5$ the surface is heated rather than the bulk. Materials with intermediate values, $10^{-2} < \varepsilon"_{eff} < 5$, are best candidates for microwave heating [8-9]. The ability of a particular material to increase its temperature when exposed to microwaves can be described by the "loss tangent" $\tan\delta = (\varepsilon"/\varepsilon')$. In most aqueous and organic media, $\tan\delta$ diminishes upon heating and microwave heating is less effective [10].

## 3. MICROWAVE HEATING

The main features which distinguish microwave heating from conventional methods are:

a) *Energy transfer*: In conventional heat treatments energy is transferred to the material through heat conduction and convection creating thermal gradients. In the case of the microwave heating, energy is directly transferred to the material through an interaction at the molecular level with the electromagnetic field (Figure 3). It has been argued that the way of energy transfer is in fact crucial in the synthesis of materials [11]. The depth that the radiation reaches varies depending on the material and other factors such as the dielectric and magnetic properties, the frequency and power of the microwave, temperature, conductivity, the size and the density of the material.

b) *Rapid heating*: Microwaves massively reduce the processing time as compared to conventional synthesis but significant qualitative changes in the crystal structure and the resulting material properties are usually not encountered. In some cases even quantitative improvements can be achieved. Some authors claim that chemical reactions, when microwave-assisted, are accelerated between 10 -1000 times compared with conventional methods [12]. A problem with such fast heating is the formation of "hot spots" and the temperature profile may not be homogeneous, which could possibly affect the reaction container [13-14].

c) *Selective heating of materials*: Microwaves can be used to carry out selective heating, which is not possible by using conventional heat treatments. Depending on the characteristics of the specific material (dielectric properties, size and structure) and its ability to couple to the electric and/or magnetic field of the microwaves, heating is located in particular areas. Since some materials do not couple well with the most frequently used 2.45 GHz microwaves at room temperature, a radiation susceptor (such as SiC, graphite or activated carbon) can be added to the precursor mixture. The susceptor strongly couples with the radiation causing a considerable increase in temperature to facilitate a more efficient microwave heating process [7, 15]. However, most polar materials do couple to the electromagnetic wave sufficiently strong to reach high temperatures, no additives are necessary and the reaction time is just a few minutes.

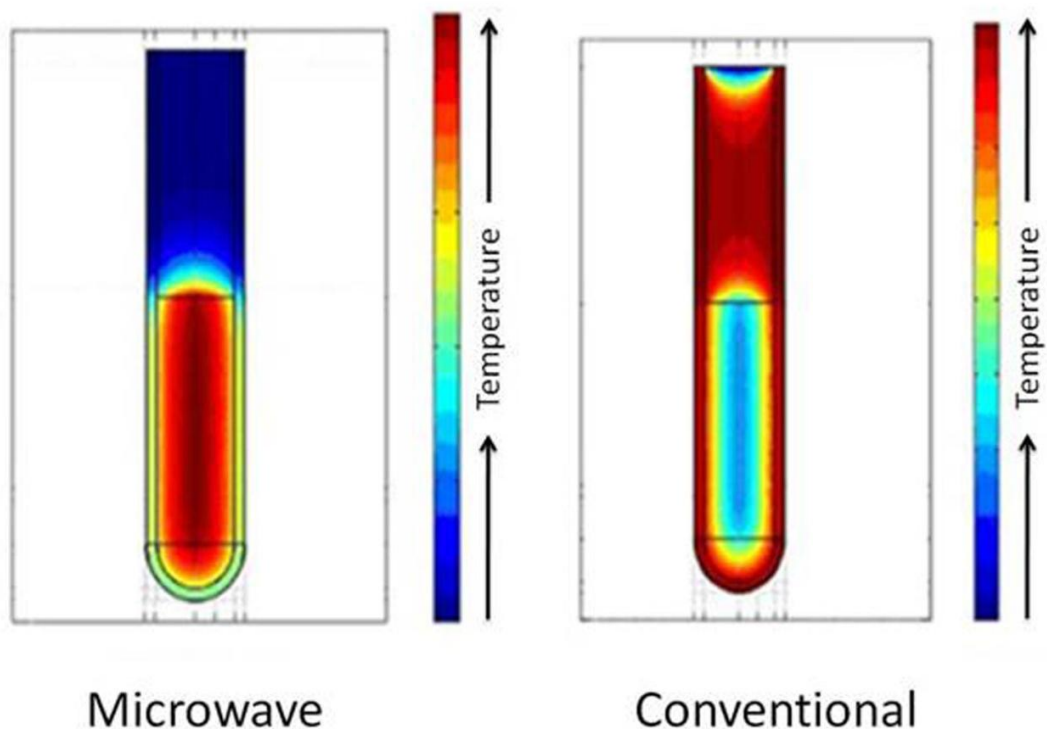

**Figure 3.** Differences in the temperature distribution in microwave and conventional processes (image downloaded from the web: http://www.biotage.com/DynPage.aspx?id=22052).

**Table 1. Maximum temperatures reached within different materials after few minutes of exposure to microwaves [8]**

| Material | t (min) | T (ºC) | Material | t (min) | T (º C) |
|---|---|---|---|---|---|
| C(amorphous) | 1 | 1283 | $V_2O_5$ | 11 | 714 |
| C(graphite) | 2 | 1100 | $Cr_2O_3$ | 7 | 130 |
| Ti | 1 | 1150 | $MnO_2$ | 6 | 1287 |
| V | 1 | 557 | $V_2O_5$ | 11 | 714 |
| Fe | 7 | 768 | $MnO_2$ | 6 | 1287 |
| Co | 3 | 697 | $Mn_2O_3$ | 6 | 1180 |
| Ni | 1 | 384 | $Fe_3O_4$ | 3 | 1258 |
| Zn | 3 | 581 | $Co_2O_3$ | 3 | 1290 |
| Zr | 6 | 462 | NiO | 6 | 1305 |
| Nb | 6 | 358 | CuO | 6 | 1012 |
| Mo | 4 | 660 | ZnO | 5 | 326 |
| W | 6 | 690 | $WO_3$ | 6 | 1270 |

In Table 1 the highest temperatures reached in selected materials are summarized.

Non-thermal effects: Many research groups are currently studying a number of anomalies involved with microwave heating that have been broadly termed "non-thermal effects", which include any differences of the microwave method as compared to conventional methods which cannot be explained by differences in the temperature and the temperature profile [16]. Today's debate focusses on anomalous effects between the electric field and the interfaces of the particles which possibly involves the formation of a plasma and the increase of the solid diffusion due to second-order effects [17-18].

## 4. MICROWAVE ASSISTED SYNTHESIS

Microwave assisted synthesis of perovskite oxides may be performed by several alternative methodologies:

i)  Direct irradiation of a mixture of the solid reactants;
ii) Irradiation of a solution in an autoclave;
iii) Combining microwave heating with other synthetic procedures such as sol-gel or combustion [4].

### 4. 1. Solid State Microwave Synthesis

There are many materials with various applications that have been synthesized by microwave irradiation of solid precursors in recent years. The most common synthetic procedure consists in mixing the precursors and packing them in a pellet, which is then deposited in an adequate crucible (usually porcelain, alumina or SiC) and placed in a microwave oven (a domestic one may be used). As mentioned above, in some cases the presence of a radiation susceptor is needed in order to generate a sufficient initial temperature for the reaction to start [19]. It has been shown that solid state microwave assisted synthesis is a very fast and effective method for the production of oxides.

#### *4. 1. 1. "Simple" $ABO_3$ Perovskites*

Stoichiometric perovskite oxides have the ideal chemical formula $ABO_3$, where A and B are cations of different sizes. A is a large cation with a similar size than the $O^{2-}$ anion. Therefore, the A-cations and $O^{2-}$ anions are cubic close-packed, whereas the smaller B-site cations reside on octahedral interstices. A-site cations are usually alkali, alkaline earth or Lanthanide rare-earth, and B is a medium-sized cation, normally a transition metal with preference for the 6-fold octahedral coordination. In our laboratory different perovskites families have been prepared by solid state microwave synthesis, namely $LaMO_3$ (M = Al, Cr, Mn, Fe, Co) [20] and $RE-CrO_3$ (RE = Rare Earth). The first family of materials, $LaMO_3$ can be used in very different technological applications such as components of solid oxide fuel cells (SOFCs), separation membranes, magneto-optical or magneto-resistant materials and substrates for thin films [21-23]. On the other hand, RE chromites have become recently the subject of much interest as possible multiferroic materials due to potential magneto-electric coupling between the $RE^{3+}$ and $Cr^{3+}$ cations [24-25].

The synthetic route for the LaMO$_3$ family is depicted in Figure 4. In the case of M = Al and Fe amorphous materials are produced in just 10 minutes. After heat treatment at 500 °C in a conventional furnace the corresponding LaAlO$_3$ and LaFeO$_3$ perovskite phases can be obtained. It should be noted that the heat treatment is done at a much lower temperature than the usual ceramic synthesis temperature ($\geq$ 1000 ° C).

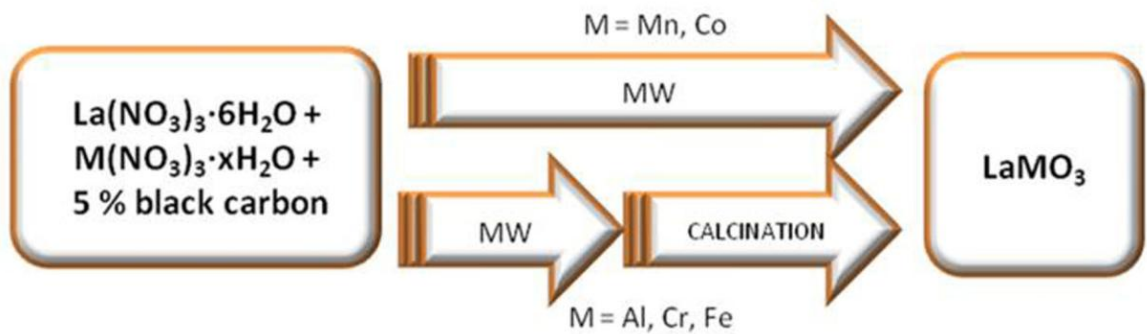

**Figure 4.** Microwave synthesis of LaMO$_3$ (M = Al, Cr, Mn, Fe, Co). Image reproduced from reference [20] with permission from Elsevier.

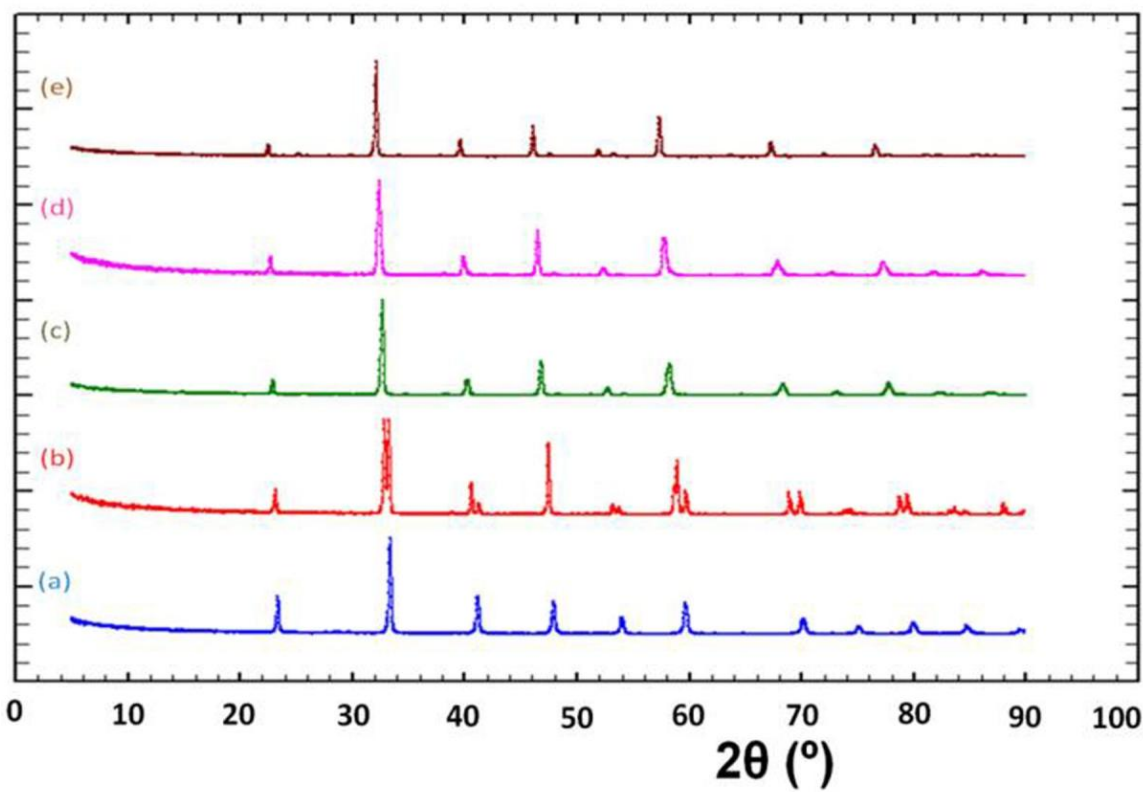

**Figure 5.** X-ray diffraction patterns of LaMO$_3$ M= (a) Al, (b) Co, (c) Cr, (d) Mn, (e) Fe.

For both $LaCoO_3$ and $LaMnO_3$, 30 minutes of microwave treatments are sufficient to obtain pure and crystalline products, without any further heating. This is due to the metallic conduction of these materials, which turns the first particles being produced into radiation susceptors and the reaction becomes self-propagated [20]. The X-ray diffraction patterns are shown in Figure 5 where the phase purity of the microwave synthesized $LaMO_3$ family is demonstrated. Figure 6 shows a scanning electron micrograph (SEM) of the $LaCoO_3$ powder, where agglomeration of particles of different sizes can be clearly seen.

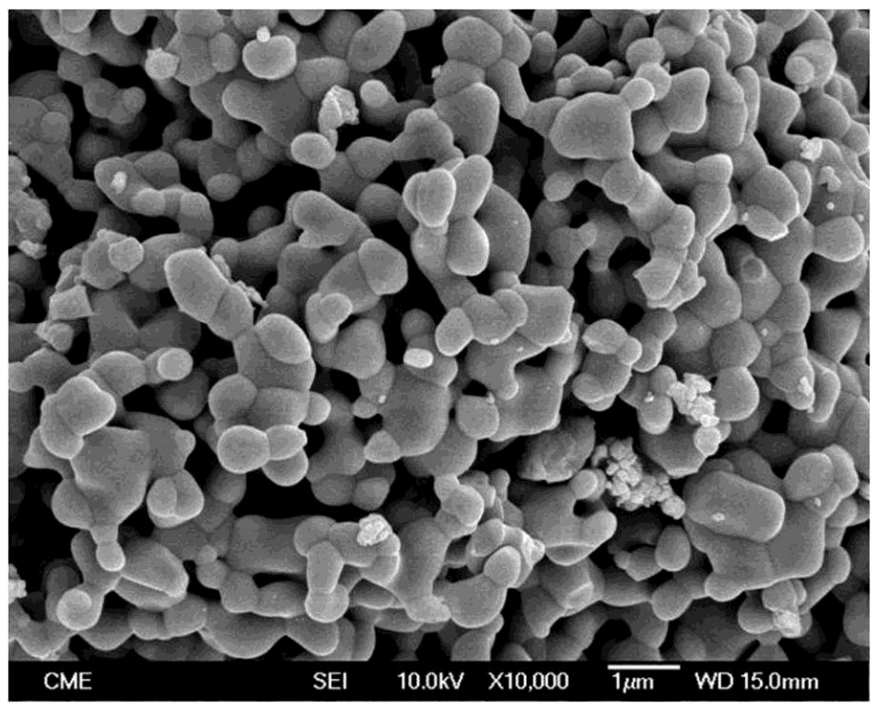

**Figure 6.** Scanning electron microscopy (SEM) image of $LaCoO_3$.

The excellent crystallinity of these materials enables to perform an accurate Rietveld refinement adjusting both the cell parameters and the atomic positions for each one of the synthesized materials. In Figure 7 the magnetic properties of microwave assisted and conventionally synthesized $LaCoO_3$ are compared. It can be seen that the curves of magnetization vs temperature are qualitatively very similar, only small quantitative differences exist mainly at low temperature as demonstrated in the Figure 7 insets.

Such differences may be ascribed to magnetic defects, which are well known to dominate the low temperature magnetism of $LaCoO_3$. The qualitative agreement of the magnetism is a clear indication that microwave assisted synthesis can successfully reproduce the $LaCoO_3$ crystal structure and the corresponding magnetic interactions between $Co^{3+}$ cations. The broad peak at ≈ 100 Kelvin displayed in the magnetization curves is commonly understood in terms of a spin-state transition, where the $Co^{3+}$ spins undergo a transformation from a low-temperature low-spin to an intermediate- or high-spin state at higher temperatures.

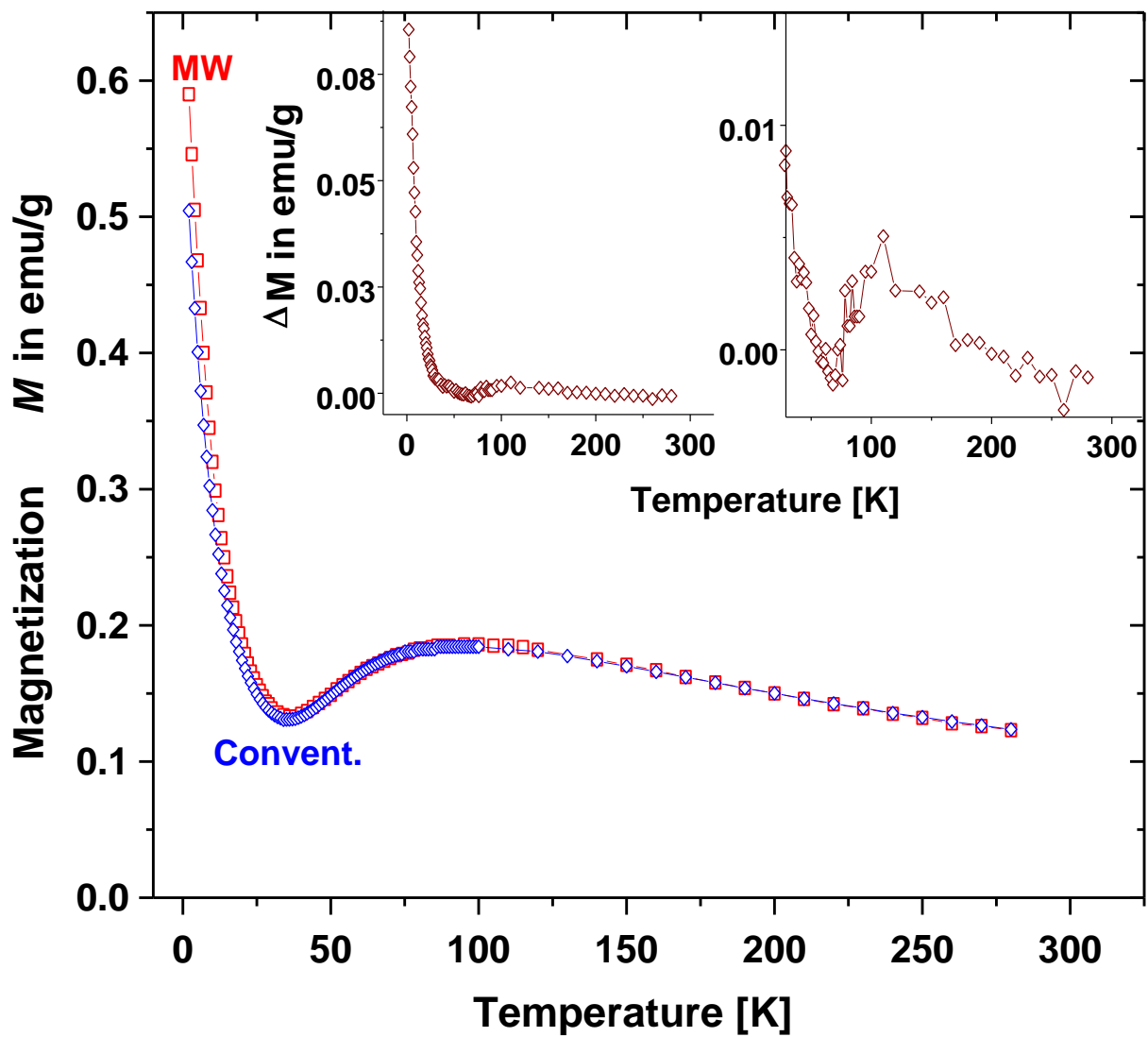

**Figure 7.** Magnetization vs temperature curves for microwave (MW) and conventionally (Convent.) synthesized $LaCoO_3$. In the Figure insets the differences in magnetization (ΔM) between the two curves are displayed at various temperature ranges.

Solid state microwave synthesis was also applied to the RE chromite series. As a representative example Figure 8a shows the Rietveld refinement performed for $ErCrO_3$ powder (space group Pbnm). The SEM micrograph (Figure 8b) of $ErCrO_3$ powder shows the agglomeration of large particles with spherical form which are made up of smaller particles near the nano-range [26]. For the full RE chromite series amorphous materials are formed during MW irradiation, which crystallize in the zircon structure $RE-CrO_4$ at 500 ºC. Upon increasing the temperature to 800 ºC the materials transform into the corresponding perovskites $RE-CrO_3$ within 2 hours [26].

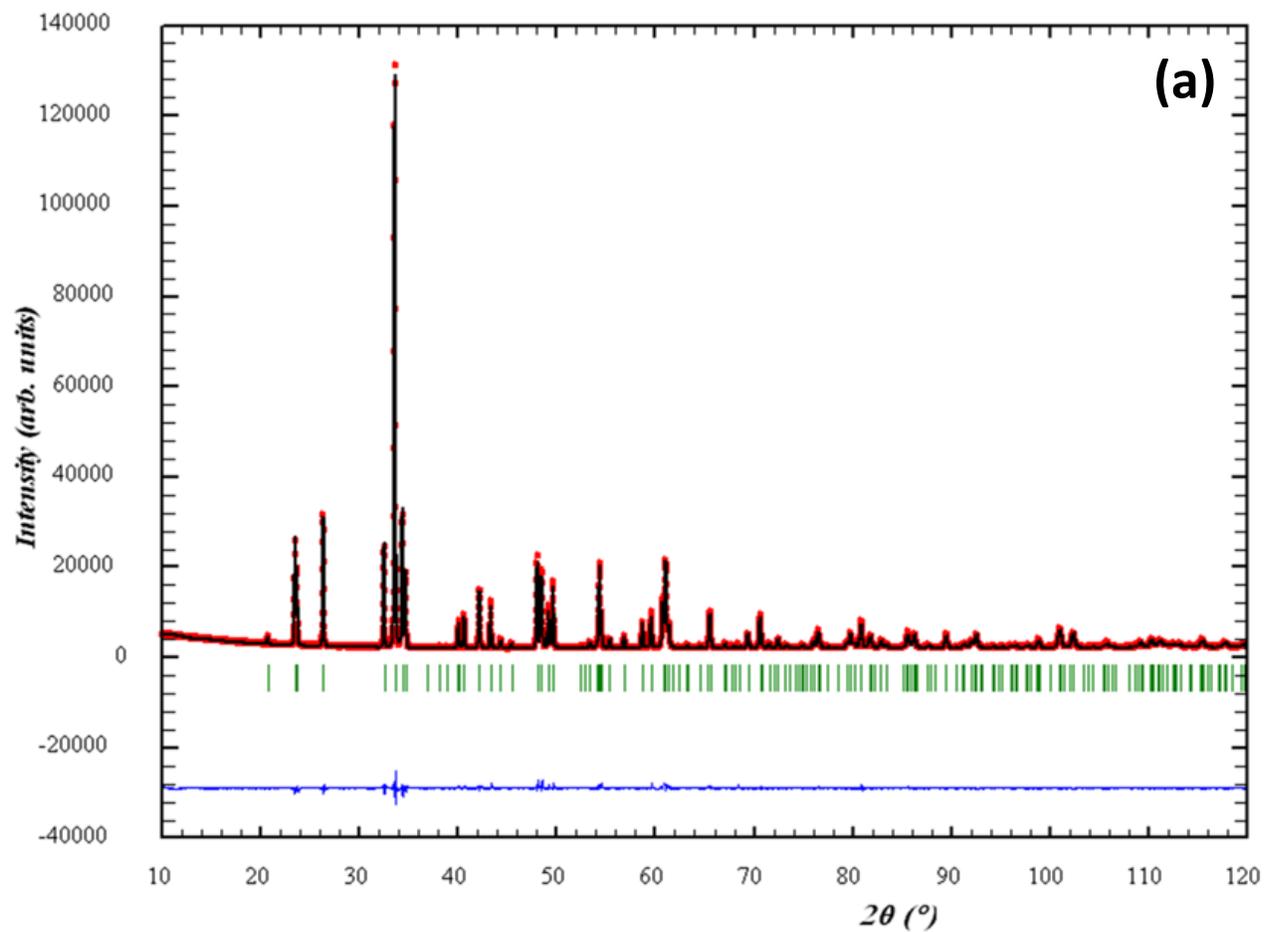

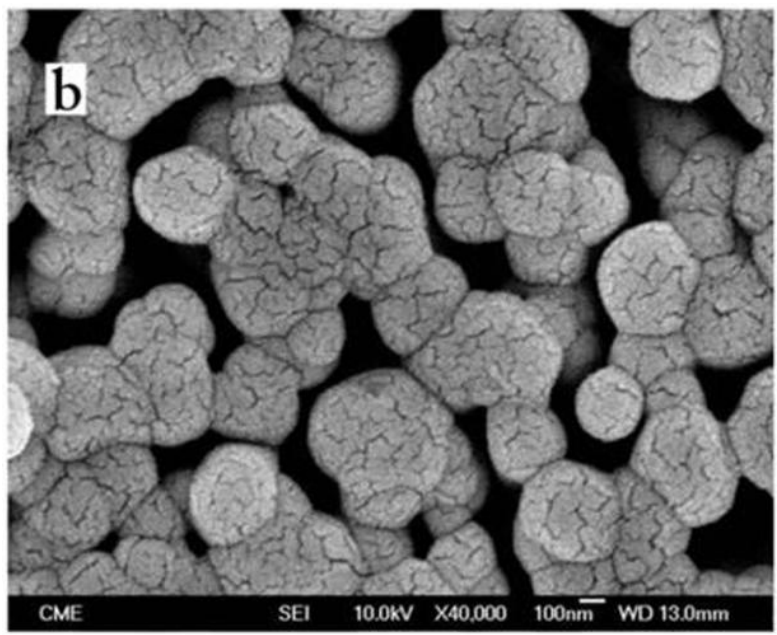

**Figure 8. (a)** Rietveld refinement of $ErCrO_3$; Space group Pbnm. Cell parameters: $a$ = 5.21389(4) Å, $b$ = 5.50302(4) Å and $c$ = 7.50391(6) Å; $R_{exp}$ = 1.74 **(b)** Scanning electron microscopy image of $ErCrO_3$.

**Figure 9.** XRD patterns of microwave synthesized (RE)CrO$_3$ powders.

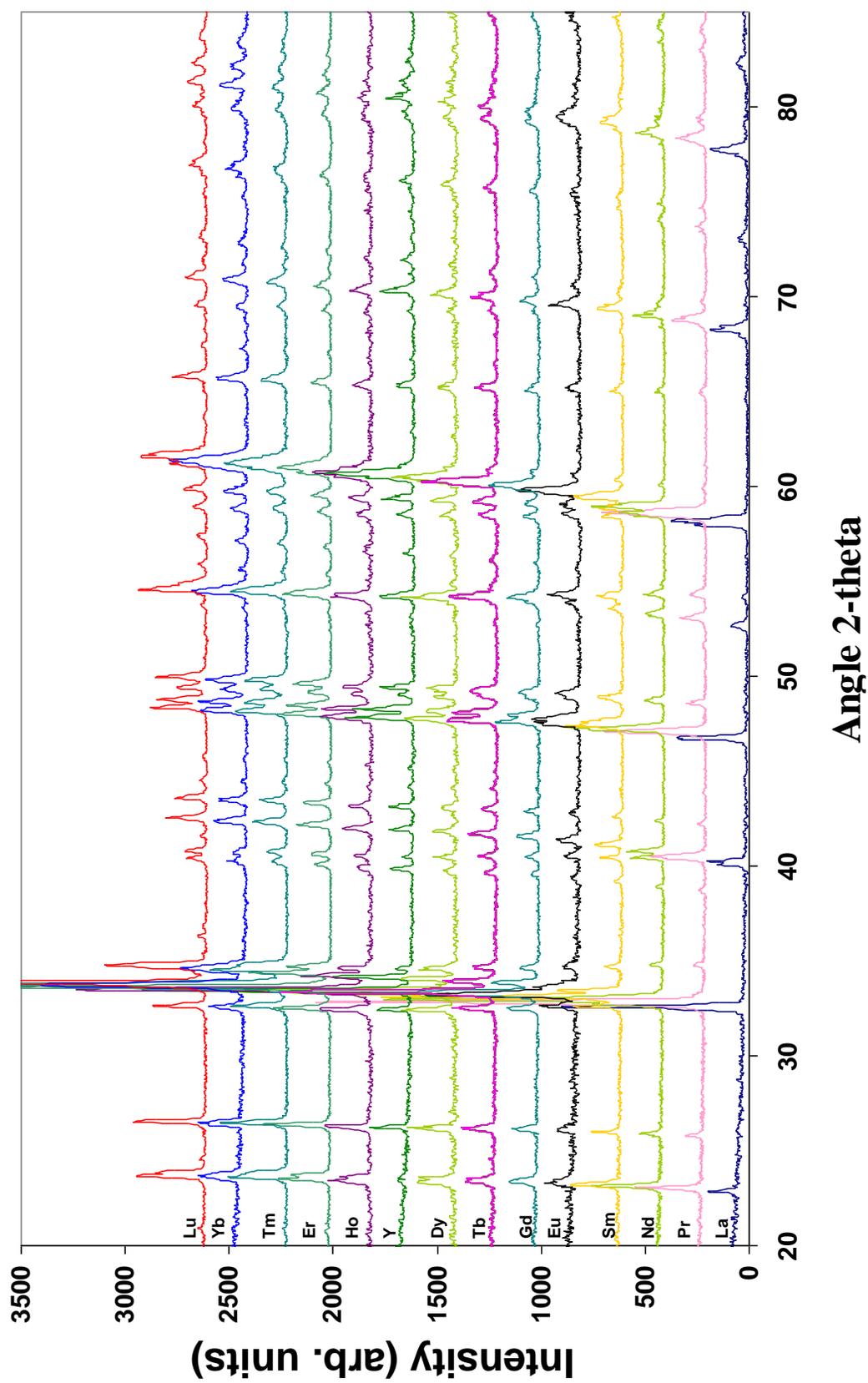

Figure 9 shows a summary of the X-ray patterns of the microwave synthesized RE chromite series, where phase purity is indicated in all compositions. The X-ray peak positions move to higher angles with decreasing RE ionic radius and increasing atomic number, which is a reflection of the decreasing unit cell size. The decrease in the ionic radius also leads to increasing octahedral tilting in the perovskite cell and decreasing Neel temperature for $Cr^{3+}$-$Cr^{3+}$ antiferromagnetic ordering, which occurs in all species.

A wealth of further perovskite structures can be formed by solid state microwave synthesis, where a selected few ones are summarized in Table 2. Vaidhyanathan et al. [27] showed a decade ago that the synthesis of some niobates and titanates, among them ferroelectrics, piezoelectrics and insulators) can be performed by microwaves in a controlled fashion.

**Table 2. Synthesis conditions for some niobates and titanates by microwave or conventional routes**

| Material | Reactants | Microwave | Conventional |
|---|---|---|---|
| $LiNbO_3$ | $Li_2CO_3$, $Nb_2O_5$ | 15 minutes 800 W | 12 h 500ºC |
| $NaNbO_3$ | $Na_2CO_3$, $Nb_2O_5$ | 17 minutes 800 W | t>>>, grindings 1250ºC |
| $KNbO_3$ | $K_2CO_3$, $Nb_2O_5$ | 12 minutes 800 W | 30 h 1000ºC |
| $BaTiO_3$ | $BaCO_3$, $TiO_2$ | 25 minutes 1000 W | t>>>, grindings 1400º C |
| $PbTiO_3$ | $PbNO_3$, $TiO_2$ | 9 minutes 600 W | 8 h 360ºC |

*4. 1. 2. Complex Perovskites*

The term "complex perovskites" is used here in reference to compositions that are more complicated than the $ABO_3$ archetype such as quaternary oxides and doped perovskites, or more complex structures such as the Ruddlesden-Popper series $A_{n+1}B_nO_{3n+1}$ or Sillen-Aurivillius phases. Perovskite superstructures are found when different species become ordered (either in the A or B sites and/or oxygen vacancies), double perovskites or some cuprate superconductors providing good examples for that behavior. It can be stated generally that the more complex the perovskites are the more difficult it is to prepare them by conventional or by microwave assisted synthesis. In more complex perovskites better diffusion processes are required to uniformly disperse 3 or more cations across the sample during the synthesis. At the current stage this still poses a major challenge for the solid state chemist; nevertheless there are several cases worth mentioning of successful microwave synthesis of complex perovskites.

### 4. 1. 2. 1. Cuprate Superconductors

In 1986 A. Müller and G. Bednorz discovered superconductivity above 30 K in the Ba-La-Cu-O system [28], which has triggered a large surge of research into these high temperature superconductors (HTSC) with many of them being cuprates with perovskite-related structures. $YBa_2Cu_3O_7$ (YBCO) may be one of the most relevant examples, because it is the first HTSC material with a superconducting transition temperature (90 K) above the boiling point of liquid nitrogen. Shortly after its discovery, the microwave-assisted synthesis of YBCO was reported [29] and the microstructure and transport properties of YBCO zone melted samples processed in a microwave cavity have been studied [30]. Nevertheless, the literature on microwave-assisted synthesis of cuprate superconductors is scarce.

Here, the microwave synthesis of superconducting $La_2CuO_4$ in a "single-step" process is presented using solid-state microwave assisted synthesis (as described previously). It is found that a microwave power threshold exists to successfully synthesize the phase pure compound at 800 W (Figure 10a). The resulting powders exhibit particle sizes in the range of 100 - 300 nm (Figure 10b). The superconducting transition temperature in this compound was found to be at ≈ 30 K upon further oxidation [31].

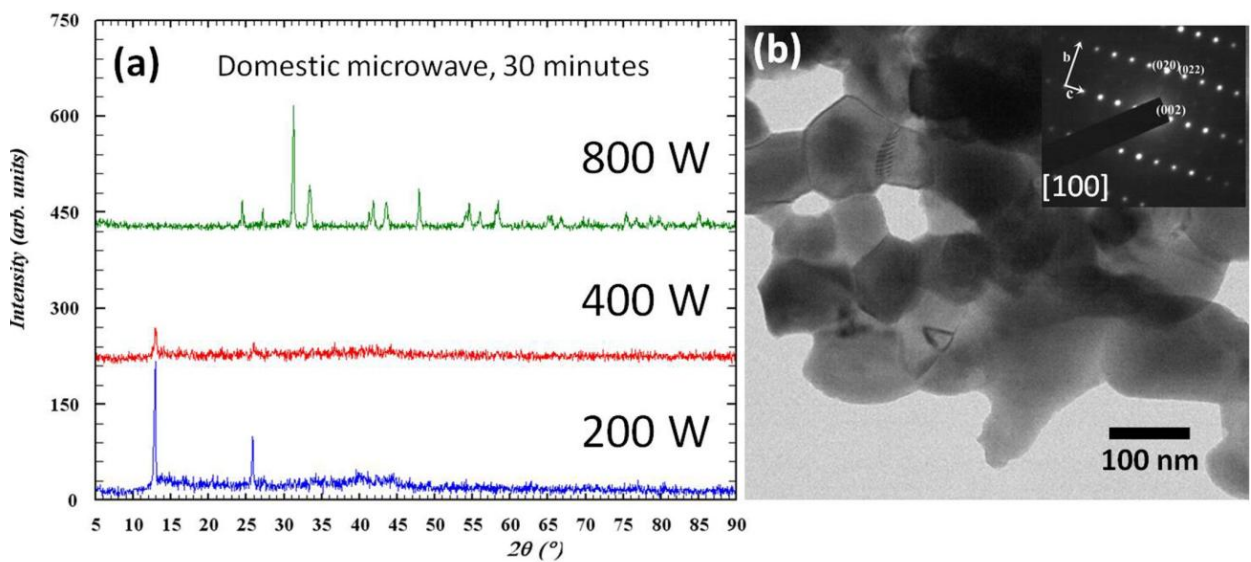

**Figure 10.** (**a**) XRD patterns of microwave synthesized $La_2CuO_4$ using different microwave power. The pure phase is obtained when using 800 W. (**b**) TEM and SAED picture (inset) of $La_2CuO_4$ sample after heating with 800 W power for 30 minutes.

### 4. 1. 2. 2. $CaCu_3Ti_4O_{12}$

$CaCu_3Ti_4O_{12}$ (CCTO) is a further complex perovskite material that can be produced by solid state microwaves synthesis. This 1:3 A-site ordered perovskite related structure is strongly tilted and is commonly indexed with space group Im-3.

It has recently attracted much attention due to its giant dielectric permittivity, which may be interesting for capacitor applications in microelectronic devices. It is quite well established though that the giant permittivity is not intrinsic but is debited to a core-shell structure of conducting bulk and insulating grain boundaries, thus forming an Internal barrier layer (IBLC) structure [32-34]. It has been reported that solid state microwave synthesis reduces the processing time and, more importantly, the IBLC structure is preserved and even higher values of the giant dielectric permittivity can be obtained [35-36]. Furthermore, it is possible to achieve higher densities in pellets and lower dielectric losses by microwave sintering.

Figure 11a shows the X-ray pattern of microwave synthesized CCTO oxide powders at different temperature for 30 min. The formation of single phase CCTO is indicated for heating at 800 °C and diffraction peaks are quite sharp indicating that the powder is well crystallized. The dielectric properties of ceramics obtained from two kinds of such microwave synthesized CCTO powders were investigated. After ceramic sintering at 1100 °C for 3 h, the relative dielectric permittivity from microwave (MS) powder (≈ 21400, at 1 kHz) is higher than that from conventional (CS) powder (≈ 10240, at 1 kHz) at room temperature as shown in Figure 11b. This can be attributed to a larger grain size of the ceramics obtained from MS powder as compared to CS [35].

a)

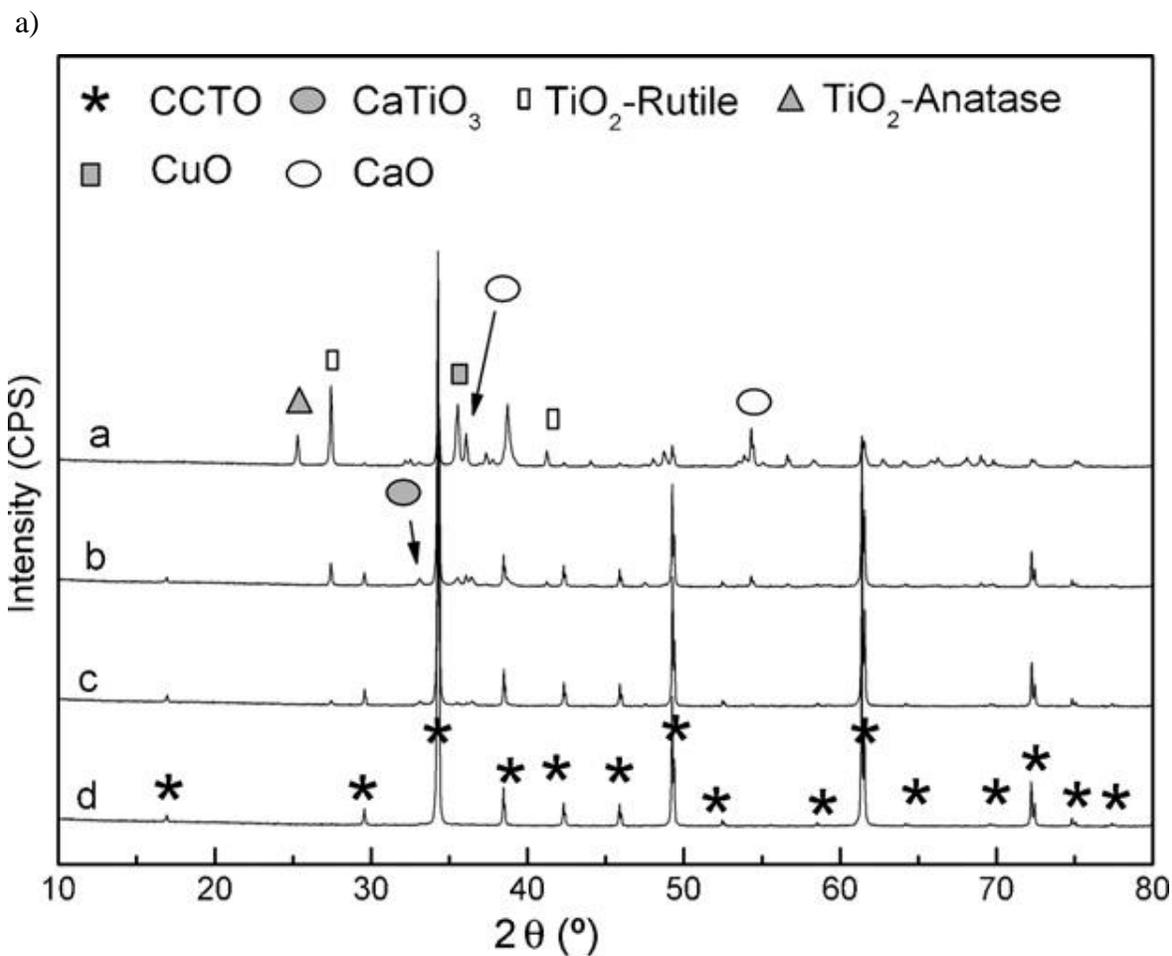

b)

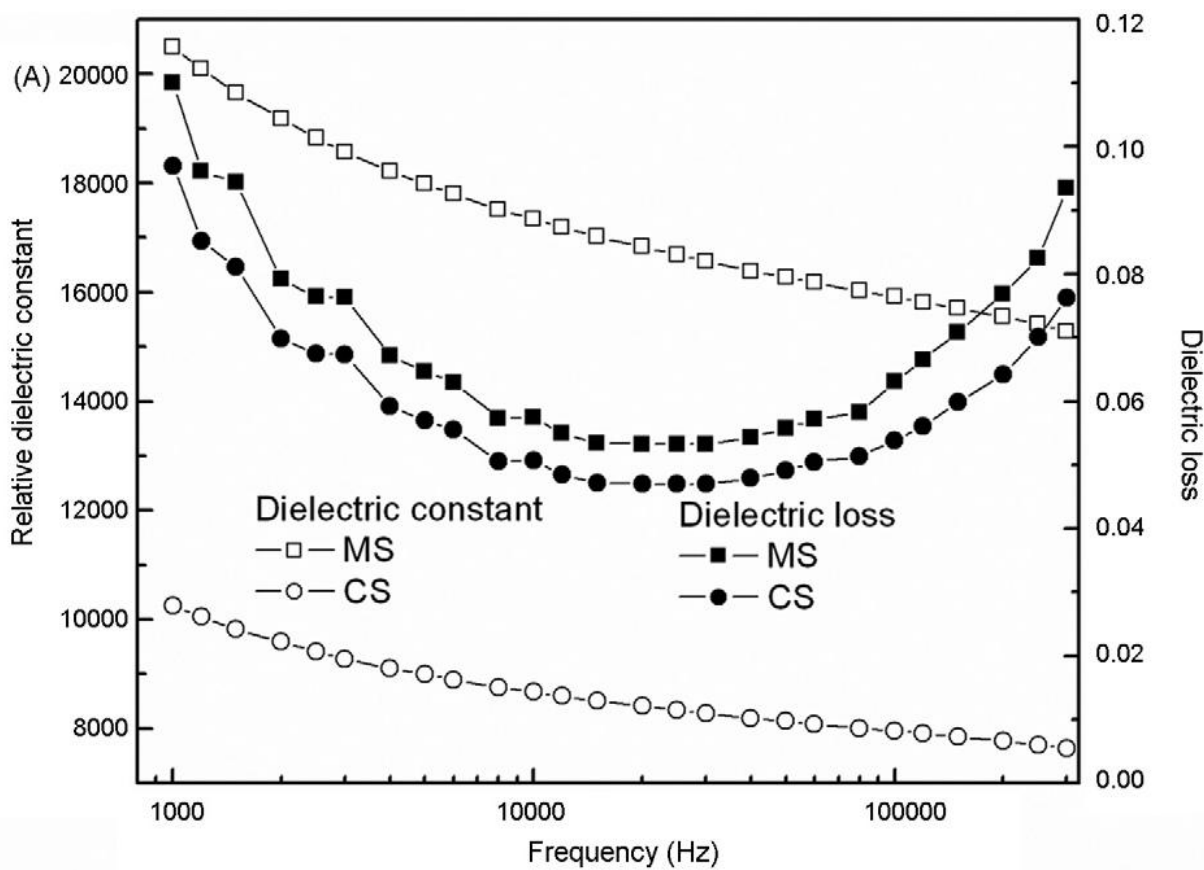

**Figure 11. (a)** XRD patterns of CCTO powder synthesized at different temperature for 30 min by microwave heating: (a) 750 C; (b) first, 800 °C; (c) second, 800 °C; (d) third, 800 °C. **(b)** Dielectric properties of sintered pellets from microwave and conventional powder. Image reproduced from reference [35] with permission from Elsevier.

## 4. 2. Single Mode Solid State Microwave-Assisted Synthesis

Conventional microwave ovens constitute a multi-mode cavity configuration, where the electric (E) and magnetic (H) components of the microwave cannot be separated. Contrarily, in a single-mode cavity the E and H distributions of the polarized microwave are utilized separately by carefully choosing the position of the sample during synthesis with respect to the maxima and minima in the amplitude of E and H (Figure 12).

The single-mode microwave equipment used in this study is based on a microwave generator working at the standard 2.45 GHz frequency with a variable power up to 2 kW and a TE10p microwave cavity. The experimental set-up includes a microwave generator, circulator (water cooled magnetron), 3-stub tuner (impedance agreement accord) and the cavity. Tuning the length of the cavity –moving a coupling iris and a short circuit piston- it is possible to work in a TE102 or TE103 resonant mode (Figure 12b). In the TE102 mode the sample is placed at the maximum amplitude of the magnetic field component of the microwave (H mode), whereas in the TE103 mode the sample is placed at the maximum of the E field (E mode). During the synthesis process the sample temperature can be measured with a pyrometer [37], which consists in a non-contacting device that intercepts and measures thermal radiation.

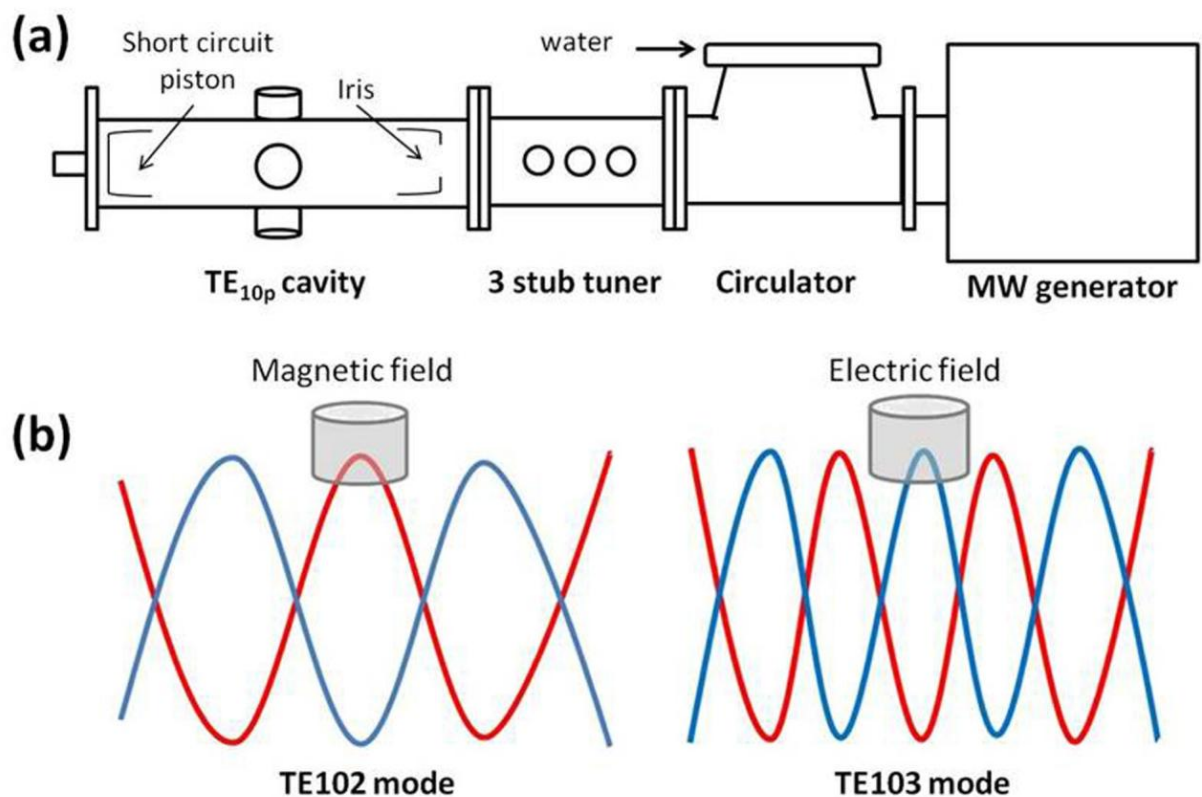

**Figure 12. (a)** Single-mode microwave apparatus. **(b)** Magnetic (red) and perpendicular electric (blue) field amplitude profiles in the cavity depending on the excited mode (TE102 or TE103). The cylinder represents the ideal positioning of the sample for each mode.

M. Gupta et al. [15] argued that the electric loss is dominant for dielectric materials (e.g. pure oxide ceramics) and the material is heated significantly only when placed in the E field, but displays modest or almost no heating when placed in the H field. However, for certain semiconducting materials and metals, the magnetic loss can be significant and the material can be heated in the magnetic field. For mixed systems containing more than one material the heating mode depends on the constituents added. If ceramic and metal are combined, sample heating takes place in both electric and magnetic fields.

### 4. 2. 1. Synthesis of Cubic BaTiO$_3$

$BaTiO_3$ is a well known ferroelectric material with high dielectric permittivity ($\varepsilon_r > 1000$) at room temperature and is one of the most widely used ceramic materials in the electronics industry, especially in multilayer ceramic capacitors. Nano-sized cubic $BaTiO_3$ powder was prepared rapidly at 90 °C using $TiO_2$ and $Ba(OH)_2$ precursors by a single-mode microwave-assisted synthesis system. $BaTiO_3$ particles of 30 to 50 nm were obtained after microwave treatment for 5 min at 90 °C (Figure 13). Conventional heating would consume approximately 7 times more energy than the microwave-assisted process at the same temperature [38].

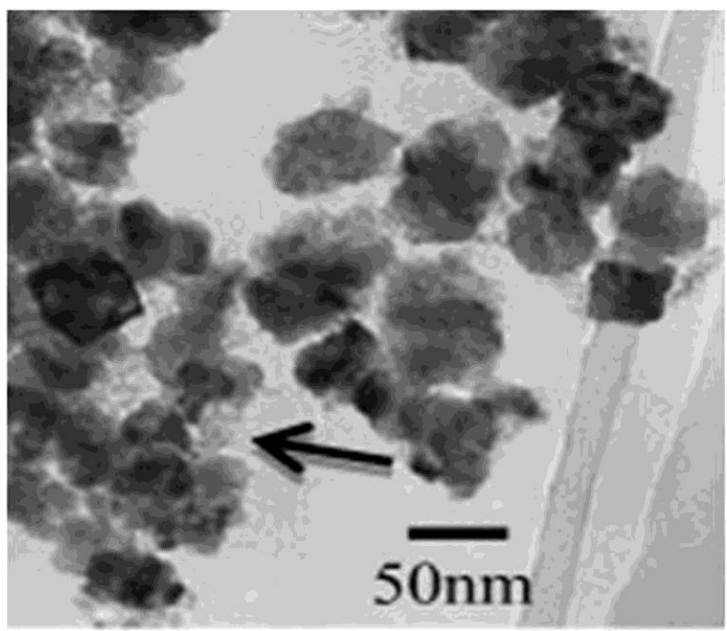

**Figure 13.** Morphologies of c-BaTiO$_3$ particles prepared at 90 °C for 5 min by single-mode microwave heating. Image reproduced from reference [38] with permission from Elsevier.

### *4. 2. 2. Synthesis of Lanthanum Chromite*

LaCrO$_3$ perovskite was synthesized in our laboratories by a single mode microwave process (SAIREM) in the TE102 mode (H mode) in just 2 minutes with a microwave power of 150 W, where the temperature was controlled to be about 900 ºC. As mentioned above, the H mode may be less common for ceramic materials but is feasible here due to the Co$^{3+}$ magnetic structure providing sufficient magnetic losses. Figure 14a shows the temperature profile of the synthesis process.

a)

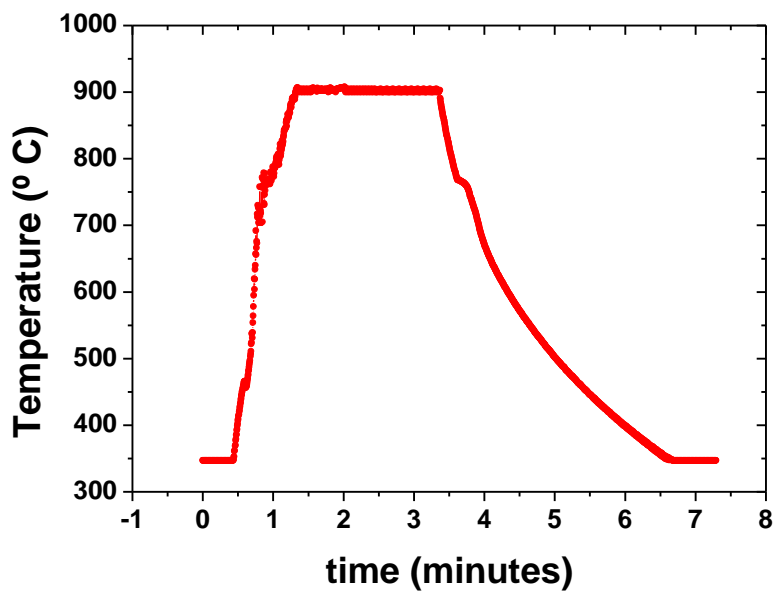

b)

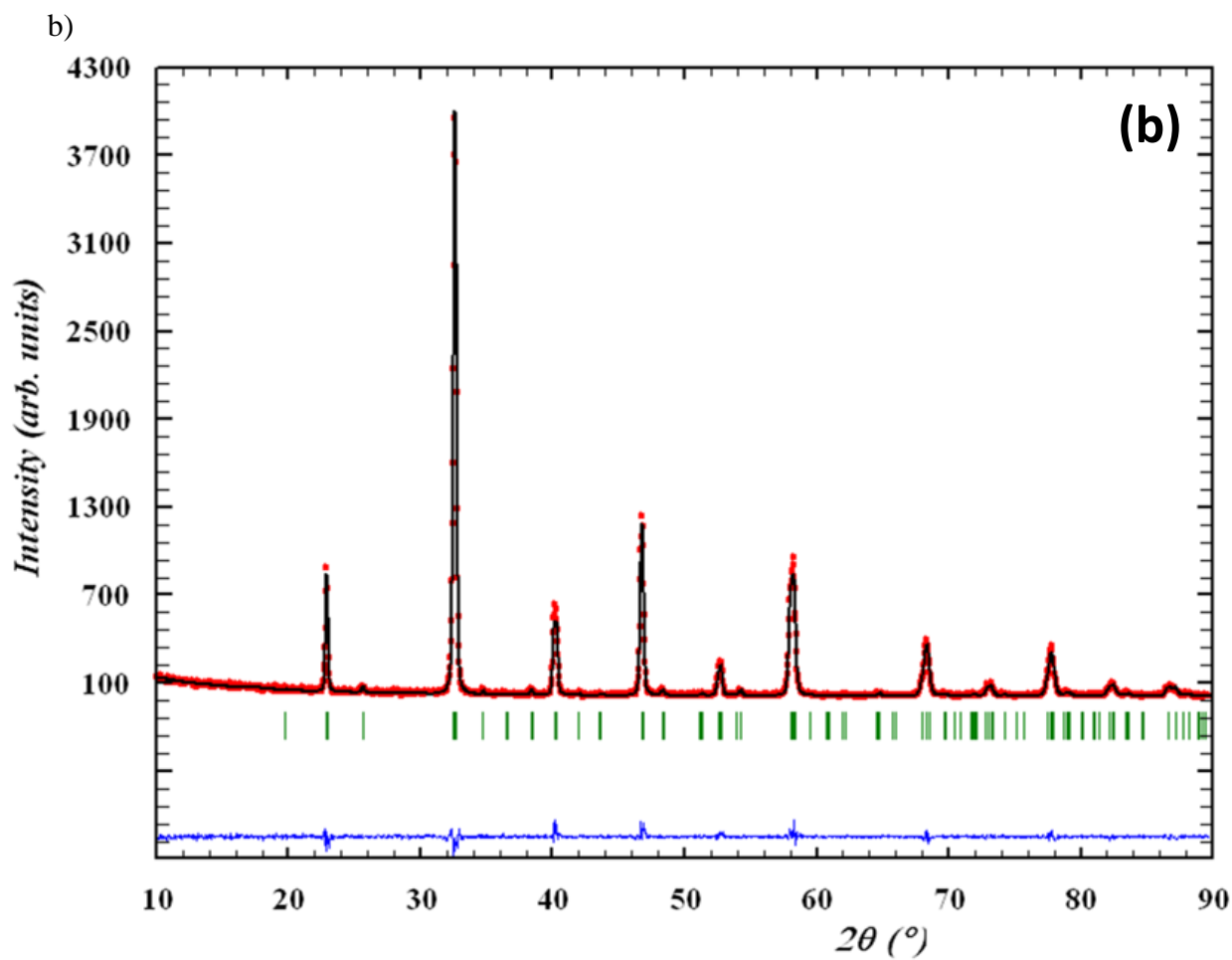

**Figure 14. (a)** Temperature control of the $LaCrO_3$ synthesis process using a pyrometer. The pure phase can be obtained in 2 minutes synthesis. **(b)** Rietveld refinement of powder X-ray diffraction patterns of $LaCrO_3$: observed (red dotted lines), refined (black solid lines), and their difference (blue bottom line). Green vertical bars indicate the X-ray reflection positions.

The starting precursors used were $La(NO_3)_3 \cdot 6H_2O$ and $Cr(NO_3)_3 \cdot 9H_2O$ in the appropriate stoichiometric ratio. The space group of the phase pure $LaCrO_3$ obtained was the expected orthorhombic Pbnm (#62) and the cell parameters from Rietveld refinement (Figure 14b) were ***a*** = 5.5092(2) Å, ***b*** = 5.4731(2) Å and ***c*** = 7.7544(3) Å.

Similarly, D. Grossin et al. [39] demonstrated single mode microwave synthesis of $La_{0.8}Sr_{0.2}MnO_3$ using a microwave TE10p cavity, where SiC was used as a microwave susceptor to facilitate an effective synthesis process within only 10 minutes. The properties of the microwave ceramic were equivalent to the properties obtained conventionally.

### 4. 3. Microwave Assisted / Hydrothermal Synthesis

In the early 1990s, Sridhar Komarneni at the University of Pennsylvania launched a pioneering work by studying and comparing the differences between hydrothermal syntheses performed with conventional means of heating and hydrothermal syntheses performed by heating special autoclaves with microwaves [40]. Since then many functional materials have been produced by microwave assisted hydrothermal synthesis, ranging from binary metallic oxides, oxyhydroxides, ternary oxides to more complex materials and structures such as zeolites or other mesoporous materials.

"Hydrothermal synthesis" is a "wet", moderate-pressure method of synthesis, which consists in an aqueous solution that is heated above 100 ºC in a sealed container (autoclave) and concomitantly the pressure increases. This results in an increase of the dispersion of the system components which react very quickly. It should be further noted that "solvothermal" synthesis has become increasingly popular lately, where non-aqueous solvents are used.

For the hydrothermal synthesis it is important to consider "subcritical" and "supercritical" synthesis conditions, where temperature is below or above the critical temperature of water ($T_c$ = 374 ºC) respectively. Above this critical point the pressure in the synthesis autoclave raises very rapidly with increasing temperature, which makes the reaction process difficult to control. Besides the characteristic microwave parameters of power and time, the method of microwave-assisted hydrothermal synthesis involves additional parameters such as the media reaction, the pH of the solution, the temperature and pressure. Therefore, more advanced synthesis technology is required to control pressure and temperature in the subcritical region of water. Autoclaves are made out of Teflon, in which case the temperature is limited to 250 ºC, or out of quartz. It should be noted that microwave assisted hydrothermal synthesis is an effective method for the production of nanoparticles, because the particle size can be controlled by tuning the mechanisms of nucleation and growth kinetics through the appropriate choice of the synthesis parameters. In the following, microwave-assisted hydrothermal synthesis of selected perovskite oxides will be presented.

### *4. 3. 1. BiFeO$_3$*

BiFeO$_3$ exhibits multiferroic properties at room temperature (ferroelectric and magnetic order coexisting in the same phase) and is, therefore, a good candidate for potential multiferroic application in information technology. BiFeO$_3$ can be synthesized in crystalline form by combined microwave hydrothermal synthesis in less than 5 minutes [41-42]. Starting reactants are Fe(NO$_3$)$_3$·9H$_2$O and Bi(NO$_3$)$_3$·5H$_2$O together with 8M KOH. Microwave assisted hydrothermal synthesis was carried out at a pressure of 15 bars and a microwave power of 500W.

The X-ray pattern presented in Figure 15a indicates phase pure BiFeO$_3$ (trigonal, R3c), and successful Rietveld refinement was performed. The unit cell parameters obtained were *a* = 5. 5799 (1) Å and *c* = 13. 8692 (4) Å. A detailed micro-structural study of the materials has been performed by means of HRTEM (Figure 15b) confirming the structural information obtained by X-ray diffraction. In Figure 16 plots of magnetization vs applied magnetic field plots are presented taken at 5 K. A small hysteretic effect may be indicative of the BiFeO$_3$ typical weak ferrimagnetic moment.

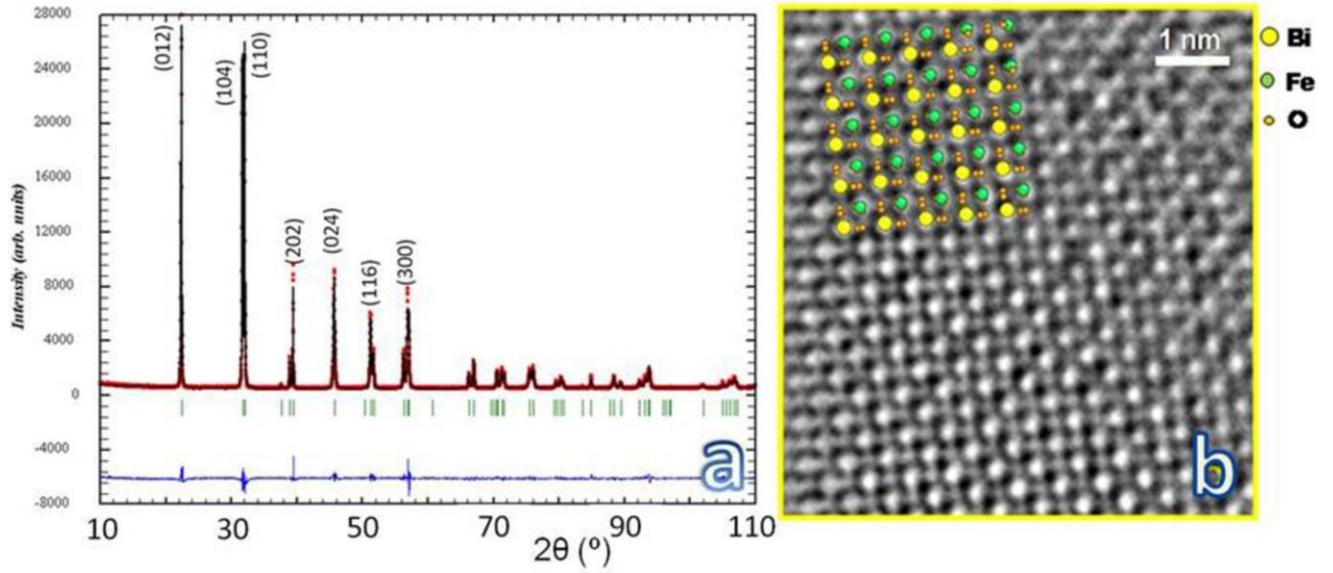

**Figure 15.** **(a)** Rietveld refinement of $BiFeO_3$ **(b)** HRTEM image of $BiFeO_3$.

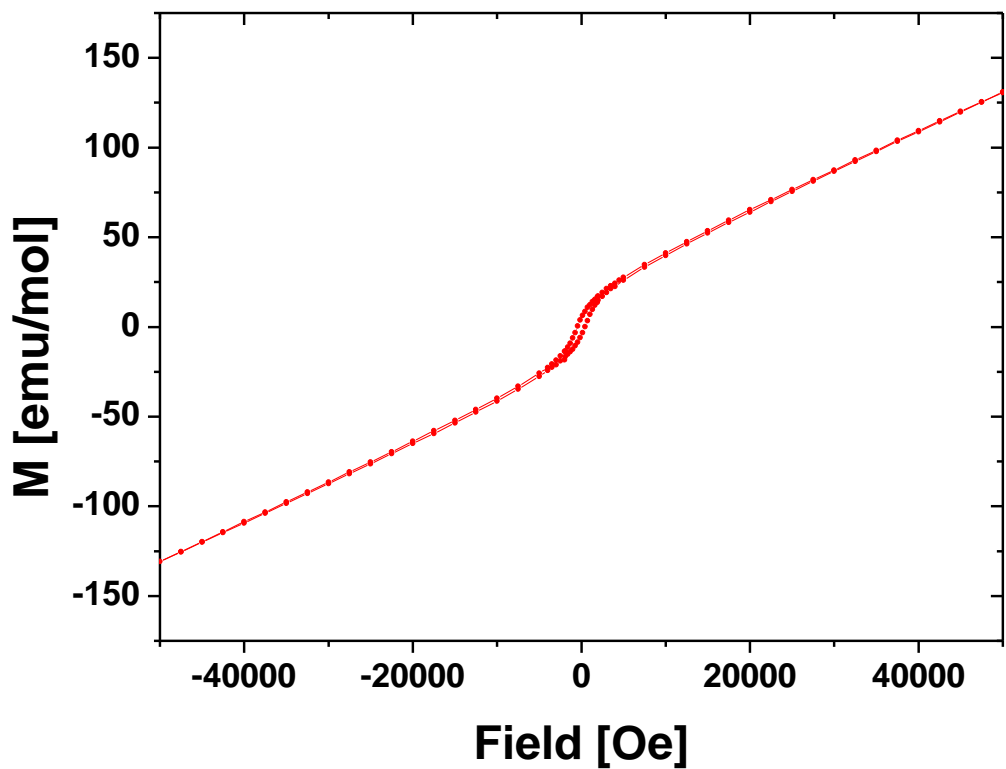

**Figure 16.** Magnetization vs applied magnetic field hysteresis loop of $BiFeO_3$ at 5 K.

### *4. 3. 2. BaTiO₃*

Novel synthesis techniques for BaTiO$_3$ have raised considerable interest recently to control the shape and size of BaTiO$_3$ particles to improve ferroelectric properties for applications.

BaTiO$_3$ perovskite was synthesized by the microwave hydrothermal route [43], starting from Ba(NO$_3$)$_2$ and TiCl$_4$, using KOH as mineralizer at a temperature of 150 °C in a EMC MDS-2000 microwave set-up. Figure 17 shows the X-ray powder diffraction patterns for the microwave hydrothermal BaTiO$_3$ powders prepared for 3, 12, and 20 h. The wide-angle (20°–80° 2θ) diffraction patterns of all samples indicate the single BaTiO$_3$ phase. The insets in Figure 17 show the reflections in the 2θ range of 44°–46°, where the asymmetrical peak of the sample prepared for 3 h can be well fitted with two distinct Gaussian peaks (dotted lines) with 1:2 intensity ratio.

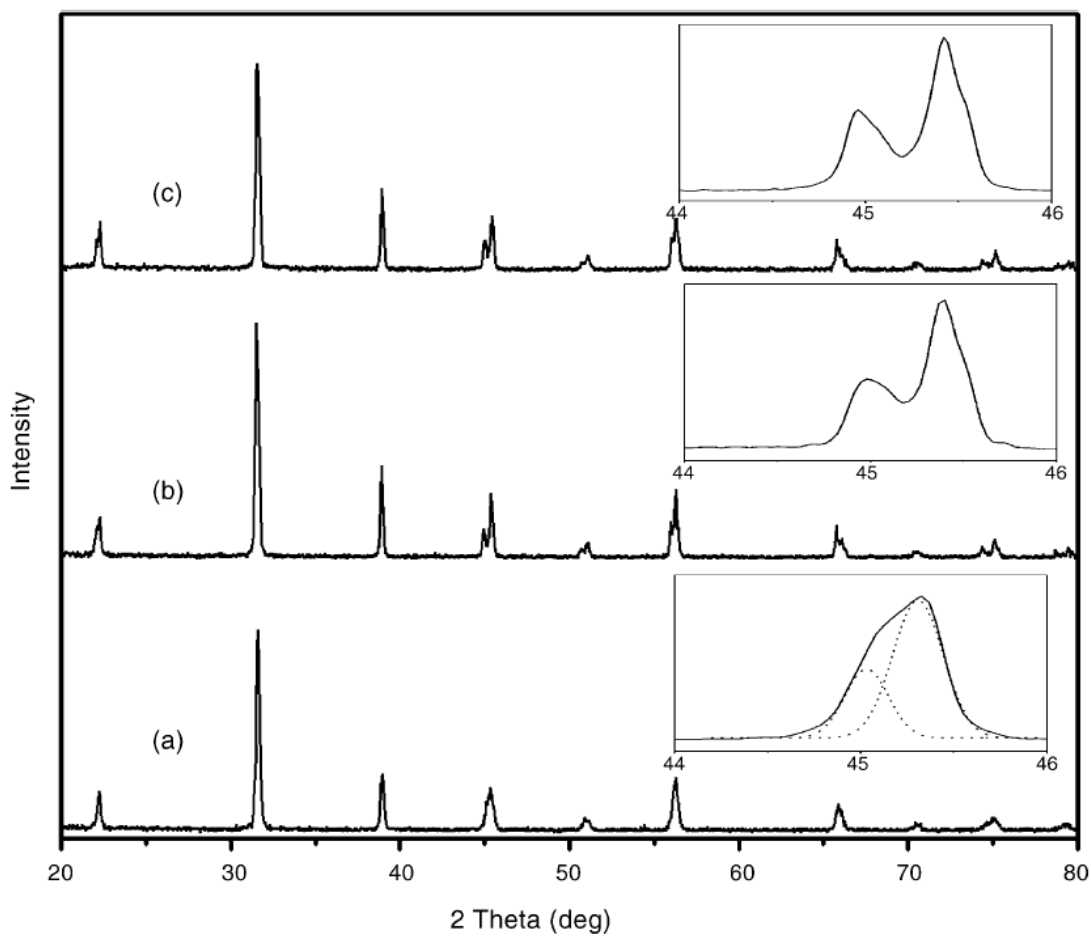

**Figure 17.** XRD patterns of BaTiO$_3$ powders prepared by microwave hydrothermal methods at 240 °C. The initial Ti concentration was 0.525 mol dm$^{-3}$ with a Ba(OH)$_2$/Ti mole ratio of 3.2. The solutions were heated for different amounts of time: **(a)** 3h. The reflection around 45° 2θ in the inset is well fitted with two Gaussian peaks (dotted lines) with an intensity ratio of 1:2, **(b)** 12 h, and **(c)** 20 h. Image reproduced from reference [43] with permission from Elsevier.

On the other hand, the reflections in the insets for the samples synthesized for 12 and 20 h display enhanced splitting. The splitting or asymmetry of reflections in this region is a result of the distortion of the unit cell, characteristic of tetragonal $BaTiO_3$ [43].

### 4. 3. Microwave Synthesis Combined with Sol-Gel or Combustion

The sol-gel process is a wet-chemical technique widely used in the synthesis of inorganic materials, typically metal oxides. The starting point is a chemical solution (sol) which acts as the precursor for a polymeric network (gel) formed upon hydrolysis and polycondensation reactions. Thermal decomposition of a particular gel (usually containing high amounts of water) leads to the corresponding oxide. This process was discovered already at the end of the 19th century, but was not fully understood until 1931 when Steven S. Kistler demonstrated that the networks in a dry gel are of the same nature than those existing in a moist one.

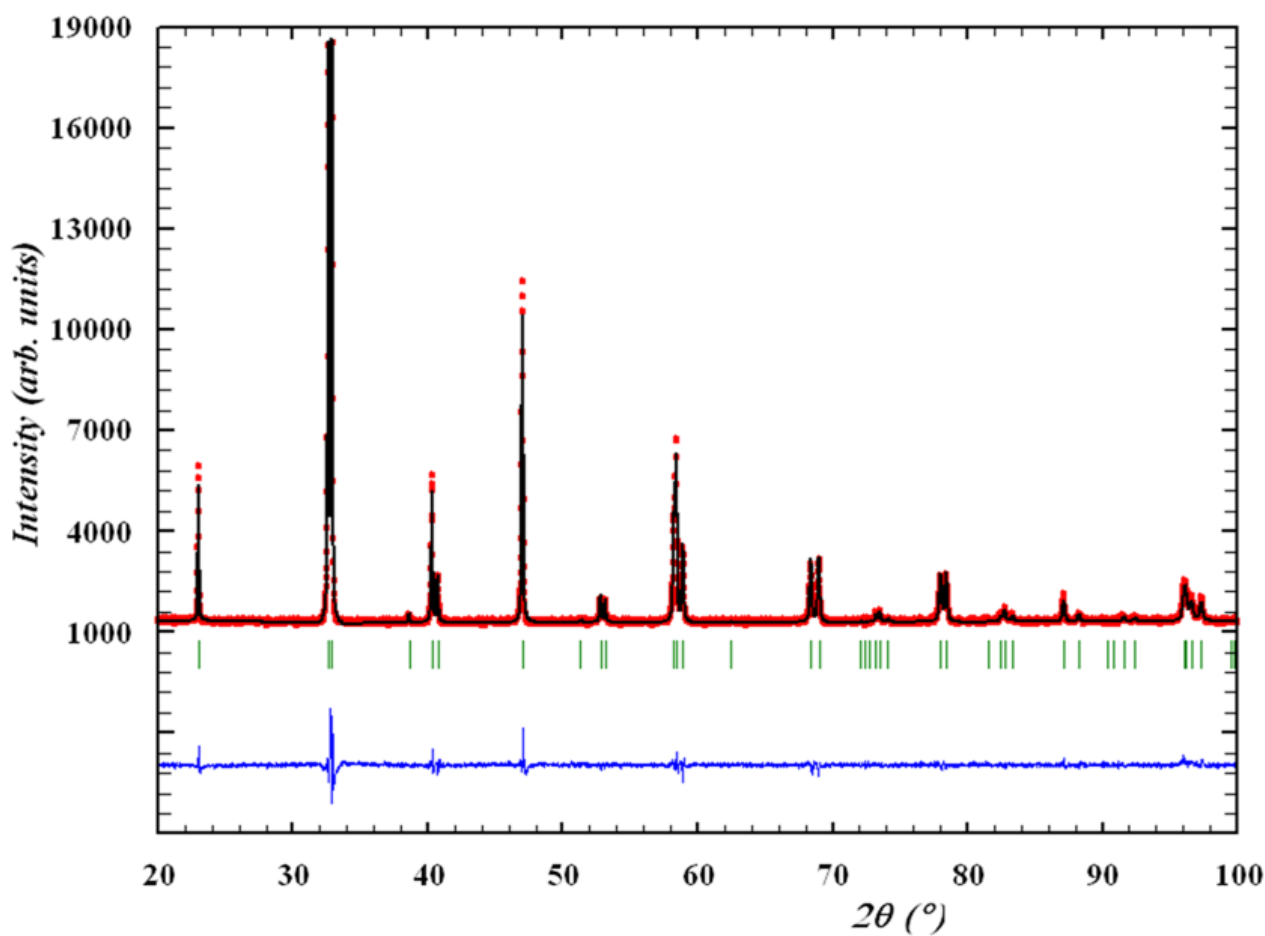

**Figure 18.** Rietveld refinement of $La_{0.8}Sr_{0.2}Fe_{0.5}Co_{0.5}O_3$; Space group R-3c. Cell parameters: $a = b = 5.4818(1)$ Å and $c = 13.2752(3)$ Å; $R_{exp} = 2.53$.

The application of microwave radiation to gels has been demonstrated for the synthesis of cathode materials for solid oxide fuel cells (SOFCs) [20], such as $La_{1-x}Sr_xFeO_{3\pm\delta}$ and $La_{1-x}Sr_xFe_yCo_{1-y}O_{3\pm\delta}$ (Figure 18).

Combustion reactions for the synthesis of oxide materials usually involve the mixing of metal nitrates and a solid fuel (urea, glycine, sacharose) in adequate quantities, where the later provides the necessary energy to ignite the (exothermic) reaction. The ignition can be performed by providing the energy in the form of microwaves. The high speed of the process together with the production of large quantities of gas produce sponge-like porous materials, many times yielding nano-sized particles. For instance, perovskite $LaMnO_3$ powders with an average crystallite size of 12.5 nm can be rapidly synthesized via a microwave-induced autocombustion reaction using glycine as a fuel and nitrate as an oxidant [44].

## 5. CONCLUSION

Microwaves are increasingly used as a new synthetic route in Solid State Chemistry. The particular nature of the matter-microwave interactions leads to rapid synthesis of many inorganic materials such as perovskite oxides. As compared to synthesis techniques where heat is transferred by convection, the reactions in microwave techniques are orders of magnitude faster. Microwave-assisted synthesis of perovskites in the solid state is often limited to "simple" compositions (ternary) but the possibility of combining microwaves with other methods such as hydrothermal synthesis, sol-gel, or combustion allows for better stoichiometric control of complex doped phases. In particular, combining microwave heating with solvothermal synthesis may result in metastable phases and novel morphologies. Single-mode polarized microwave synthesis allows separating the magnetic and electric components and provides an accurate control of the temperature together with much shorter reaction and processing times.

The synthesis of a wide range of perovskite oxide materials was shown to be feasible by microwave techniques, where such materials include superconducting, ferromagnetic, ferroelectric, dielectric and multiferroic perovskite systems.

## 6. ACKNOWLEDGEMENTS

The authors acknowledge funding from the Community of Madrid (Materyener S2009/PPQ-1626) and the Ministry of Science and Innovation (MAT-2007-64006). R.S. acknowledges the Ministerio de Ciencia e Innovación (MICINN) for granting a Ramon y Cajal Fellowship.

Many thanks for valuable contributions go to Ian Terry (Durham University, UK), Dra. C. Parada (UCM), Dra. M.E. Villafuerte-Castrejón (UNAM-Mexico), Dr. L. Fuentes (CIMAV-Mexico), Dr. D. Avila (UCM), Dr. S. Marinel (CRISMAT-CNRS, Caen) and I. Herrero-Ansorregui (UCM).

## 7. REFERENCES

[1]   Mitchell R. H. *Perovskites Modern and Ancient*, Almaz Press Inc.: Canada, 2002; Vol. 57, 317.
[2]   Hayes, B. L. *Microwave Synthesis: Chemistry at the Speed of Light*, C. E. M. publishing: 2002; 11-27.
[3]   Sutton W. H. *Ceram. Bull*. 1989, 68, 376-386.
[4]   Prado-Gonjal, J.; Morán, E. *An. Quím*. 2011, 107(2), 129–136.
[5]   Zhao, J.; Yan, W. *Modern Inorganic Synthetic Chemistry*, 2011,8, 173-195.
[6]   Clark, D. E.; Folz, D. C.; Folgar, C.; Mahmoud, M. Microwave Solutions for Ceramic Engineers, *The American Ceramics Society*, Inc.: Westerville, O. H., 2005; 5-30
[7]   Gupta, M.; Leong, W. W. *Microwaves and Metals, John Wiley and sons*: Asia, 2007, 35-60.
[8]   Rao, K. J.; Vaidhyanathan, B.; Ganduli, M.; Ramakrishnan P. A. *Chem. Mater*. 2009, 11, 882-895.
[9]   Sahu, R. K.; Rao, M. L.; Manoharan, S. S. *Journal of Materials Science* 2001, 36, 4099-4102.
[10]  Lidström, P.; Tierney, J.; Wathey, B.; Westman, *J. Tetrahedron*, 2001, 57, 9225-9283.
[11]  Balaji, S.; Mutharasu, D.; Subramanian, N. S.; Ramanathan, K. *Ionics*, 2009, 15, 765-777.
[12]  Parada, C.; Morán, E. *Chem. Mat*. 2006, 18, 2719-2725.
[13]  Liu, S.-F.; Abothu, I. R.; Komarneni, S. *Mater. Let*. 1999, 38, 344-350.
[14]  Komarneni; S.; Katsuki, H. *Pure Appli. Chem*. 2002, 9, 1537-1543.
[15]  Das, S.; Mukhopadhyay, A. K.; Datta, S.; Basu, D. Bull. *Mater. Sci*., 2009, 32, 1–13.
[16]  Herrero, M. A.; Kremsner, J. M.; Kappe, C. O.; *J. Org. Chem*. 2008, 73, 36-47.
[17]  Wroe, R.; Rowley, A. T. J. *Mater. Sci*. 1996, 31, 2019-2026.
[18]  De la Hoz, A.; Díaz-Ortiz, A.; Moreno, A. *Chem. Soc. Rev*. 2005, 34,164-178.
[19]  Chandrasekaran S., Basak T., Ramanathan S., *J. Mater. Process. Techn*. 2011, 211, 482-487.
[20]  Prado-Gonjal, J.; Arévalo-López, A. M.; Morán, E. *Mat. Res. Bull*. 2011, 46, 222-230.
[21]  Backhaus-Ricoult, M. *Solid State Sci*. 2008, 10, 670-688.
[22]  Choi, Y.; Mebane, D. S.; Lin, M. C.; Liu, M. *Chem. Mater*. 2007, 19, 1690-1699.
[23]  Petrovic, S.; Terlecki-Baricevic, A.; Karanovic, L.; Kirilov-Stefanov, P.; Zdujic, M.; Dondur, V.; Paneva, D.; Mitov, I.; Rakic, V. *Applied Catalysis B:Environmental* 2008, 79, 186-198.
[24]  Sahu, J. R.; Serrao, C. R.; Ray, N.; Waghmare, U. V.; Rao, C. N. R., *J. Mater. Chem*. 2007, 17, 42-44.
[25]  Serrao, C. R.; Kundu, A. K.; Krupanidhi, S. B.; Waghmare, U. V.; Rao, C. N. R. *Phys. Rev. B*. 2005, 72, 220101–220104.
[26]  Prado-Gonjal, J.; Schmidt, R.; Ávila, D.; Amador, U.; Morán, E., *J. Eur. Ceram. Soc.* 2012, 32, 611–618.
[27]  Vaidhyanathan, B.; Raizada, P.; Rao, K. J., *J. Mater. Sci. let*. 1997, 16, 2022-2025.
[28]  Bednorz, J. G.; Müller, A. *Zeitschrift für Physik B Condensed Matter* 1986, 64, 189-193.
[29]  Baghurst, D. R.; Chippindale, A. M.; Mingos, D. M. P. *Nature*, 1988, 322, 24.
[30]  Marinel, S.; Bourgault, D.; Belmont, O.; Sotelo, A.; Desgardin, G. *Physica C*, 1999, 315, 205-214.
[31]  Herrero-Ansorregui, I.; Prado-Gonjal, J.; Morán, E., U. C. M. master thesis, 2012 (unpublished work).
[32]  Chung, S.-Y.; Kim, I.-D.; Kang S.-J. *Nature Materials* 2004, 3, 774-778.

[33]  Homes, C.; Vogt, T.; Shapiro, S. M.; Wakimoto, S.; Ramirez, A. P. *Science* 2001, 293,673-676.
[34]  Schmidt, R.; Stennett, M. C.; Hyatt, N. C.; Pokorny, J.; Prado-Gonjal, J.; Li, M.; Sinclair, D. C., *J. Eur. Ceram. Soc*. 2012, 32, 3313–3323.
[35]  Yu, H.; Liu, H.; Luo, D.; Cao, M., *J. Mater. Proces. Tech*. 2008, 208, 145-148.

[36] Hatagalung, S. D.; Ibrahim, M. I. M.; Ahmad, Z. A. *Ceramics International* 2008, 34, 939-942.
[37] Marinel, S.; Savary, E.; Gomina, M., *J. M. P. E. E.* 2010, 44, 57-63.
[38] Katsuki, H.; Furuta, S.; Komarneni, S. *Mater. Lett.* 2012, 83, 8–10.
[39] Grossin, D.; Marinel, S.; Noudem, J.-G. *Ceramics International*, 2006, 32, 911-915.
[40] Komarneni, S.; Roy, R.; Li, Q. H. *Mat. Res. Bull.* 1992, 27, 1393-1405.
[41] Prado-Gonjal, J.; Villafuerte-Castrejón, M. E.; Fuentes, L.; Morán, E. *Mat. Res. Bull* 2009, 44, 1734-1737.
[42] Prado-Gonjal, J.; Ávila, D.; Villafuerte-Castrejón, M. E.; González-García, F.; Fuentes, L.; Gómez, R. W., Pérez-Mazariego, J. L.; Marquina, V.; Morán, E. *Solid State Sci*. 2011, 13, 2030-2036.
[43] Sun, W.; Li, C.; Li, J; Liu, W. *Mat. Chem. Phys*. 2006, 97, 481–487.
[44] Weifan, C.; Fengsheng, L.; Leili, L.; Yang, L. *J. Rare Earths*, 2006, 24, 782–787.